\documentclass[pra,aps,nofootinbib,hyperref,onecolumn,showpacs,preprint]{revtex4}

\usepackage{graphics}
\usepackage{epsfig}
\usepackage{natbib}
\usepackage{amssymb}
\usepackage{bm,amsmath}

\begin{document}
\title{Doppler cooling of three-level $\Lambda$-systems by coherent pulse trains}

\author{Ekaterina Ilinova}
\affiliation{Department of Physics, University of Nevada, Reno,
Nevada 89557, USA}

\author{Andrei Derevianko}
\affiliation{Department of Physics, University of Nevada, Reno,
Nevada 89557, USA}

\begin{abstract}
We explore the possibility of decelerating and Doppler cooling an ensemble of tree-level $\Lambda$-type atoms by a coherent train of short, non-overlapping laser pulses. We show that $\Lambda$-atoms can be Doppler cooled without additional repumping of the population from the intermediate ground state. We derive analytical expression for the scattering force in the quasi-steady-state regime and analyze its dependence on pulse train parameters. Based on this analysis we propose a method of choosing pulse train  parameters to optimize the cooling process.
\end{abstract}

\maketitle
\section{Introduction}
Doppler cooling \cite{HanSch75} relies
on radiative force originating from momentum transfer to atoms  from a laser field and subsequent spontaneous emission in random directions.
Cooling by CW lasers has been widely studied both theoretically and experimentally within the last several decades \cite{MinLet87Book,MetStr99Book,BerMal10_Book}.
Schemes for cooling the two-level atoms by the trains of ultrashort laser pulses \cite{Hof88, Str89,WatOhmTan96,Kie06, IliAhmDer11} were proposed. 
 The possibility of stimulated cooling the two-level atoms by the pairs of counter propagating $\pi$-pulses \cite{Kaz74,NoeNoeSch96,GoeBloHau97, SodGriOvc97} and similar idea of cooling by bichromatic standing wave \cite{SodGriOvc97} were studied both theoretically and experimentally. The interest in cooling by the pulse trains is  stimulated by the rapid development of a pulsed laser technology and frequency combs (FC) \cite{SchHarYos08,AdlCosTho09,VodSorSor11}. In particular the mechanical action of FC on atoms was observed experimentally  in Ref.~\cite{MarStoLaw04}. 

In many cases the atom can not be approximated as a two-level system because the excited state may decay to some intermediate sublevels. As an example, group III atoms have no single-frequency closed transition on which the cooling of the ground state could be based, because their ground states are composed of two fine-structure sublevels $nP_{1/2}$ and $nP_{3/2}$.
CW laser cooling of this type of $\Lambda$-systems
in the presence of bichromatic force-assisted velocity-selective coherent population trapping has been studied in \cite{PruAri03}. Other schemes of CW sub-Doppler cooling of three-level atoms based on velocity-selective coherent population trapping have been proposed earlier \cite{AspAriKai88,KasChu92}.  There were also proposals for bichromatic force cooling of three-level $\Lambda$-atoms \cite{GupXiePad93,AspDalHei86}.

Here we propose the scheme for decelerating and cooling the three-level atoms with
the ultrafast pulse train.  In our scheme both ground states of the $\Lambda$-type system are coupled to the excited state by the same laser field.  As a result, the cooling does not require additional repumping of  population from the intermediate state.
The exerted scattering force depends on atomic velocity via the Doppler shift.  Similar to the case of a  two-level system, studied in \cite{IliAhmDer11}, the spectral profile of the scattering force mimics the periodic structure of the frequency comb (FC) spectra.
Since the positions of FC teeth depend on the pulse-to-pulse carrier envelope phase offset, CEPO, the
velocity-dependence of the scattering force can be varied in time by simply changing the phase offset between subsequent pulses. Thereby, continuous compression of velocity distribution in velocity space can be achieved.
During the pulse-train cooling, continuous velocity distributions gravitate toward a series of sharp peaks (typically of the Doppler width) in the velocity space, reflecting the underlying frequency comb structure.

There are several motivations for this work.
Wide spectral coverage of FC allows one to cool the atoms in a broad range of velocities at the same time.
In some cases, FC cooling  could be used for reducing number of required lasers.
Cooling setup based on tunable FC can be alternative to Zeeman slowers , whose fields may be detrimental for precision measurements~\cite{ZhuOatHal91}.
The presented analysis is applicable for laser cooling in ion storage rings \cite{SchKleBoo90etal,MieGriGri96etal} where the circulating  ions  are subjected to
 chopped laser field.

We start consideration by  deriving analytical expression for the scattering force in the quasi-steady-state regime (QSS), based on the expression for the density matrix obtained in our previous work \cite{IliDer12}.
In the quasi-steady-state regime the radiative decay-induced drop in the excited state population between two pulses is fully restored by the second  pulse.
This regime is similar to the saturation regime in a classical system of two kicked coupled damped oscillators.
Based on our analytical expressions, we show that the $\Lambda$-system can be Doppler cooled without additional repumping of population from the intermediate ground state.
We analyze the dependence of the scattering force on the FC  parameters. Based on this analysis we propose a principle of choosing FC parameters for optimal cooling of ensemble of $\Lambda$-type three-level atoms.

For the pulse-train-driven $\Lambda$-system there are two major qualitative effects: ``memory'' and ``pathway-interference'' effects.
Both effects play an important role in understanding of the radiative force exerted by the pulse train on the multilevel system.
The system retains the memory of the preceding pulse as long as the population of the excited state does not completely decay between subsequent pulses. This is satisfied for finite values of the product $\gamma T$, $\gamma$ being the excited state radiative decay rate and $T$ being the pulse repetition period.  Then the quantum-mechanical amplitudes driven by successive pulses interfere and the response of the system reflects the underlying frequency-comb structure of the pulse train. If we fix the atomic lifetime and increase the period between the pulses, the interference pattern is expected to ``wash out'', with a complete loss of memory in the limit $\gamma T \gg 1$.  This memory effect is qualitatively identical to the case of the two-level system, explored in Ref.~\cite{IliAhmDer11}.

The ``pathway-interference'' effect is unique for  multilevel systems. The excited-state amplitude arises from simultaneous excitations of the two ground states. The two excitation pathways interfere. The ``pathway-interference'' effect is perhaps most dramatic  in the coherent population trapping (CPT) regime \cite{Har97,MorVia11,SoaAra07,SoaAra10}  where the ``dark'' superposition of the ground states conspires to interfere destructively, so that there is no population transfer to the excited state at all.

This paper is organized as follows. In Section II we derive analytical expression for the scattering force exerted on atoms by the pulse train in a quasi steady-state regime.
In Section III  we study the dependence of scattering force on FC parameters and propose a method of their optimization.
In Section IV we study the process of cooling of thermal beam of  three-level $\Lambda$-type atoms by pulse train. We demonstrate that in the optimal cooling regime the initial velocity distribution evolves to a comb-like profile with sharp equidistant maxima, ``velocity comb''. The width of each peak is determined by the Doppler temperature  limit. Finally, the conclusions are drawn in Section V.

\section{Analytical expression for a scattering force exerted on atoms by
delta-function pulse train}

\subsection{Delta-function-like pulse model}
\label{Sec:Propagators}

As in our previous work \cite{IliDer12} we parametrize the
electric field of the pulse train at a fixed spatial coordinate as
\begin{equation}
\mathbf{E}(t)=\hat{\varepsilon}\,E_{p}\,\sum\limits_{m}\cos(\omega_{c}%
t+\Phi_{m})\,g(t-mT)
\label{Eq:TrainField} \, ,
\end{equation}
where $\hat{\varepsilon}$ is the polarization vector, $E_{p}$ is
the field amplitude, and $\Phi_{m}$ is the phase shift. The
frequency $\omega_{c}$ is the carrier frequency of the laser field
and $g(t)$ is the shape of the pulses. We normalize $g(t)$ so that $\max |g( t)|
\equiv 1$, then $E_{p}$ has the meaning of the peak amplitude.
While typically pulses have identical shapes and
$\Phi_{m}=m\phi$, one may want to install an active optical
element at the output of the cavity that could vary the phase and
the shape of the pulses.

\begin{figure}[h]
\begin{center}
\includegraphics*[width=4.0in]{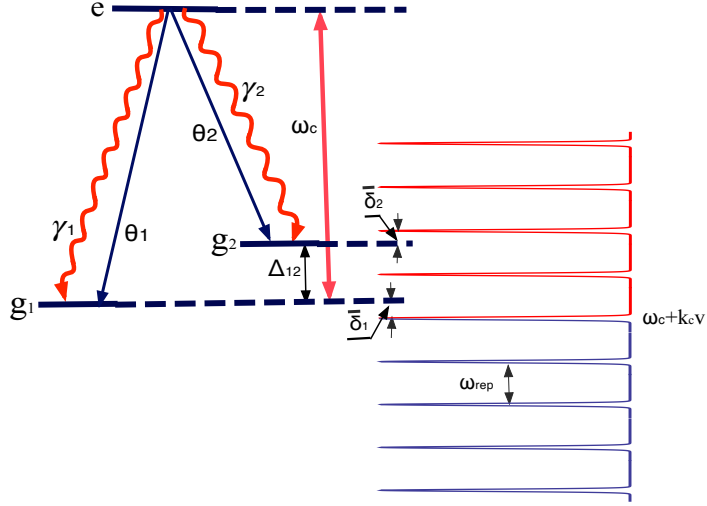}
\end{center}
\caption{ (Color online )   Energy levels of $\Lambda$-type system and positions of frequency comb teeth. The comb is Doppler shifted in the atomic frame moving with velocity $v$. \label{Fig:lambdasystem}}%
\label{Fig:Setup}%
\end{figure}

The  $\Lambda$-system, Fig.~\ref{Fig:lambdasystem}, is composed of the excited state
$|e\rangle$ and the ground states $|g_1\rangle$, $|g_2\rangle$ separated by $\Delta_{12}$; the transition frequencies between the excited and each of the ground states are $\omega_{eg_1}$, $\omega_{eg_2}$ correspondingly.
The single pulse area corresponding to a transition $|g_j\rangle\rightarrow|e\rangle$ is
\begin{equation}
\theta_j=\Omega_{j}^{peak}\,\int\limits_{-\infty}^{\infty}g(t)dt \, ,
\end{equation}
where $\Omega^{peak}_j=\frac{E_{p}%
}{\hbar}\langle e|\mathbf{D}\cdot\hat{\varepsilon}|g\rangle$ is  the peak Rabi frequency expressed in terms of the dipole matrix element.
As long as the duration of the pulse $\tau_p$ is
much shorter than the repetition time, the atomic system behaves
as if it was a subject to a perturbation by a series of
delta-function-like pulses: $\Omega^{peak}_jg(t) \rightarrow \theta_j \delta(t)$. In this limit, the only relevant
parameter affecting the quantum-mechanical time evolution is the
effective area of the pulse. The optical Bloch equations, in  rotating wave approximation, may be written in form:

\begin{eqnarray}
\dot{\rho}_{ee}&=&-\gamma\rho_{ee}-\sum\limits_{n=0}^{N-1}\delta(t-nT)\sum\limits_{j=1}^2(\theta_{j}\label{Eq:DMdeltamodel} Im\left[e^{-i(k_cz(t)-\delta_jt-\Phi_n)}\rho_{eg_j}\right],\\
\dot{\rho}_{eg_j}&=&-\frac{\gamma}2\rho_{eg_j}+\frac{i}2 \sum\limits_{n=0}^{N-1}\delta(t-nT)\sum\limits_{p=1}^{2} \theta_{p}e^{i(k_cz(t)-\delta_pt-\Phi_n)}(\rho_{ee}\delta_{jp}-\rho_{g_pg_j}), \\
\dot{\rho}_{g_jg_{j'}}&=&\delta_{jj'}\gamma_j \rho_{ee}+\frac{i}2\sum\limits_{n=0}^{N-1}\delta(t-nT)(\theta_{j'}e^{i(k_cz(t)-\delta_{j'}t-\Phi_n)}\rho_{g_je}-\theta_{j}e^{-i(k_cz(t)-\delta_jt-\Phi_n)}\rho_{eg_{j'}}),
\end{eqnarray}
where the detunings  $\delta_{j}=\omega_c-\omega_{eg_j}$  are the detunings of the carrier frequency from the frequencies of transitions $|g_j\rangle\rightarrow|e\rangle$.

The dynamics of three-level $\Lambda$-type system driven by the coherent train of delta-function like pulses
has been studied in detail in our previous work \cite{IliDer12}.
Here we employ analytical expression for the density matrix in a quasi-steady-state regime from that work. Although general expression for the density matrix was presented there, here we restrict ourself to the case most commonly realized.
If the energy gap between the two ground states is much smaller than the frequency of transition from the ground to excited state, then the ratio of
decay rates $\gamma_1/\gamma_2$ is proportional to the ratio of relevant dipole matrix elements in the same way as the ratio of pulse areas $\theta_1/\theta_2$.
In this case we can use the following parametrization:  $\theta_1/\theta_2=\gamma_1/\gamma_2=\tan\chi$.  Then the post-pulse excited state population  $\left( \rho_{ee}^s \right)_r$ in QSS regime  reads

\begin{eqnarray}
\left(\rho_{ee}^s\right)_r&=&8e^{\frac{\gamma T}2 } \sin^2\frac{\Theta}2\sin ^2\pi\kappa/D\nonumber\\
D&=&8 \cos 2\chi\left(4 \sin^4\frac{\Theta }{4}+\sin ^2\frac{\Theta }{2} \cos 2\pi\kappa\right)\sin\pi\kappa\sin \left(\bar{\eta} +\pi\kappa\right)+\nonumber\\&&
+\cos\pi\kappa\cos \left(\bar{\eta} +\pi\kappa\right)\left(4 \cos\frac{\Theta }{2}\left(\cos2\pi\kappa-5\right)+(\cos\Theta+3) (3 \cos2\pi\kappa+1)-\right.\nonumber\\&&
\left. -16 \sin ^4\frac{\Theta }{4}\sin ^2\pi\kappa\cos (4 \chi )\right)-\nonumber\\&&
-4\cosh\left(\frac{\gamma T}2\right)\left(4 \cos^2\frac{\Theta }{4}\cos2\pi\kappa+2 \cos\frac{\Theta }{2}-\cos\Theta-5\right).\label{Eq:rgen}
\end{eqnarray}

In this formula and below we employ the following notation (see also Fig.~\ref{Fig:lambdasystem})

\begin{enumerate}
\item[(i)]{ The effective single-pulse area
\begin{equation}\Theta=\sqrt{\theta_1^2+\theta_2^2},\end{equation}
where $\theta_j$ are the single-pulse areas for the two transitions $|g_j\rangle\rightarrow|e\rangle$, $j=1,2$.
}

\item[(ii)]{ Number of teeth fitting in the energy gap $\hbar\Delta_{12}$ between the two ground states
\begin{equation} \kappa=\Delta_{12}/\omega_{rep} \, .\end{equation}
Notice that $\kappa$   generally is not an integer number. When it is integer, the two-photon resonance condition is satisfied and the system evolves into the dark state.}

\item[(iii)]{ Doppler shifted phase offset between subsequent pulses
\begin{equation}
\overline{\eta}=\eta(t)-\eta(t+T)=\left(k_cv+\delta_1\right)T+\phi.
\end{equation}
Here $v$ is the atomic velocity and $\phi$ is the carrier-envelope phase offset between subsequent pulses, i.e., $\phi=\Phi_{m} - \Phi_{m+1}$ in Eq.~(\ref{Eq:TrainField}). These phase parameter will be used to characterize the spectral profile of the scattering force. As shown below the density matrix of a system and the scattering force are periodic functions of $\overline{\eta}$. }
\item[(iv)]{ Residual detunings  $\overline{\delta}_j$, $j=1,2$, between $|g_j\rangle$ levels and the nearest FC modes in the reference frame moving with the atom.
In general,
$\bar{\delta}_1=(\bar{\eta}+2\pi n_1)/T$ and  $\overline{\delta}_2=(\bar{\eta}+2\pi\kappa+2\pi n_2)/T$, where integers $n_j$ are chosen to renormalize the
residual detunings to the interval
$-\omega_{rep}/2<\overline{\delta}_j<\omega_{rep}/2$. }
\end{enumerate}
Eq.~(\ref{Eq:rgen}) gives the value of the excited state population just after the pulse. The time evolution between the pulses is described by
($mT<t<\left(m+1\right)T)$)
\begin{eqnarray}\label{decayeq}
\rho_{ee}^s(t) &=&\left(\rho_{ee}^s\right)_{r}e^{-\gamma t}.
\end{eqnarray}

The dependence on the phase offset $\bar{\eta}$ is the result of interference between the elementary responses of a system to subsequent pulses (the persistent ``memory'' of the system).
Particularly, when  $\gamma T\rightarrow\infty$, the excited state completely decays between the pulses and the interference factor vanishes (the ``memory'' is erased),

\begin{equation}
\left(\rho_{ee}^s\right)_r\rightarrow  \frac{ 4\sin^2\left(\pi\kappa\right)}{\tan^2\frac{\Theta }{4}+\frac{\sin^2\left(\pi\kappa\right)}{\sin^2\frac{\Theta }{4}}}.
\label{Eq:DMQSSbranchlargedecay}
\end{equation}

At equal pulse areas $\theta_1=\theta_2$ and decay rates  $\gamma_1=\gamma_2$ ($\chi=\pi/4$) the equation (\ref{Eq:rgen}) can be simplified further

\begin{eqnarray}
\left(\rho_{ee}^{s}\right)_r&=&\frac{e^{\frac{\gamma T}{2}} \sin ^2\left(\pi\kappa\right) \sin^2\frac{\Theta}2}{4D'}, \nonumber\\
D'&=& \left(\cos \left(\pi\kappa\right) \cos
   \left(\bar{\eta}+\pi\kappa\right) \left(\cos ^2\left(\pi\kappa\right) \cos^4\left(\frac{\Theta }{4}\right)-\cos\frac{\Theta}2
   \right)+\right.\nonumber\\
  && \left.\cosh \left(\frac{\gamma T }{2}\right) \left(\sin^4\left(\frac{\Theta }{4}\right)+\cos^2\left(\frac{\Theta }{4}\right) \sin ^2\left(\pi\kappa\right)\right)\right).   \label{Eq:DMQSSsimplea}
\end{eqnarray}

\subsection{Scattering force}
Now we focus on the evaluation of the
cooling force,
\begin{equation}\label{ClFrc}
F_z=\hbar k_c \sum\limits_{j=1}^2\mathrm{Im}[\rho_{eg_j}\Omega_{eg_j}] \, .
\end{equation}
The laser field is present only during the
pulse, so effectively we deal with a sum over instantaneous forces
\begin{equation}
\mathbf{F}(t)=p_{r}\,\sum\limits_{m,j}\theta_j\delta(t-mT)\,\mathrm{Im}%
[e^{-i(k_cz(t)-\delta_jt-\Phi(t))}\rho_{eg_j}(t)]~\mathbf{\hat{k}}_{c}, \label{Eq:Force}%
\end{equation}
where $\mathbf{\hat{k}}_{c}$ is the unit vector along the
direction of the pulse propagation. The change in the
linear momentum of a particle due to the $m$-th pulse is $\Delta\mathbf{p}%
_{m}=\lim_{\varepsilon\rightarrow0^{+}}\int_{mT-\varepsilon}^{mT+\varepsilon
}\mathbf{F}(t)dt$. We find

\begin{equation}
\frac{-\Delta\mathbf{p}_{m}}{p_{r}}=\left(  \left(
\rho_{ee}^{m}\right) _{r}-\left(  \rho_{ee}^{m}\right)
_{l}\right)  ~\mathbf{\hat{k}}_{c},
\label{Eq:DPoverP}%
\end{equation}
where $\left(\rho_{ee}^{m}\right) _{r}=\rho_{ee}(mT+\varepsilon)$,  $\left( \rho_{ee}^{m}\right)_{l}=\rho_{ee}(mT-\varepsilon)$ ($\tau\ll\varepsilon\ll T$)  are the excited state population values just before and just after the pulse.

This result follows from noticing that $\Delta p_{m}$ is an
integral of a particular combination
$\,\sum\limits_j\delta(t-mT)\theta_j\mathrm{Im}[e^{-i(k_cz(t)+\delta_jt+\Phi(t)}\rho_{eg_j}(t)]$ over time.
This combination enters the r.h.s. of Eq.~(\ref{Eq:DMdeltamodel}).
Then by integrating Eq.~(\ref{Eq:Force}) over time we
immediately arrive at Eq.~(\ref{Eq:DPoverP}).

Several insights may be gained from analyzing Eq.(\ref{Eq:DPoverP}).
\begin{enumerate}
\item[(i)]{ Eq.~(\ref{Eq:DPoverP}) simply states that the averaged over a big number of cycles single laser pulse
fractional momentum kick is equal to a difference of
populations before and after the pulse. 
}
%
%
\item[(ii)]{ As elucidated earlier for CW laser cooling (see e.g., Ref.~\cite{MetStr99Book})
radiative decay plays a crucial role in maintaining
force directed along the laser beam. In the context of the
pulse-train cooling, Eq.(\ref{Eq:DPoverP}), radiative decay brings down
the excited state population in the time-interval between the
pulses, thus keeping pre- and post-pulse excited
state population difference negative; this leads to a net force
along the direction of the pulse train propagation.
}

\item[(iii)]{ In the regime when two FC modes match both transition frequencies between the excited and ground states,
the system evolves into a ``dark'' superposition of two ground states which is transparent to the pulses.
The population of the excited state in this case and consequently the scattering force are both zero.
}
\end{enumerate}
In the quasi-steady-state regime the value of single pulse fractional momentum kick is
\begin{equation}
\frac{-\Delta p_{s}}{p_{r}}=\left(\rho_{ee}^{s}\right)
_{r}~\times\left(
1-e^{-\gamma T}\right) \, \label{DPoverPQ}%
\end{equation}
and the average scattering force can be represented as
\begin{equation}
F_{sc}=\frac{\Delta p_s}{T}.\label{Eq:forcegeneral}
\end{equation}

In a particular case of equal branching ratios $b_1=b_2=1/2$, the expression for the scattering force reads (this was obtained using Eq. (\ref{Eq:DMQSSsimplea}) )

\begin{eqnarray}\label{Eq:scforce}
F_{sc}&=&\frac{\Delta p}{T}=-\frac{\hbar k_c}{ T}\frac{\sinh{\gamma T /2} \sin ^2\left(\frac{\Theta }{2}\right) \sin ^2(\pi  \kappa )}{D'}\end{eqnarray}

\begin{figure}[h]
\begin{center}
\includegraphics*[width=3.2in]{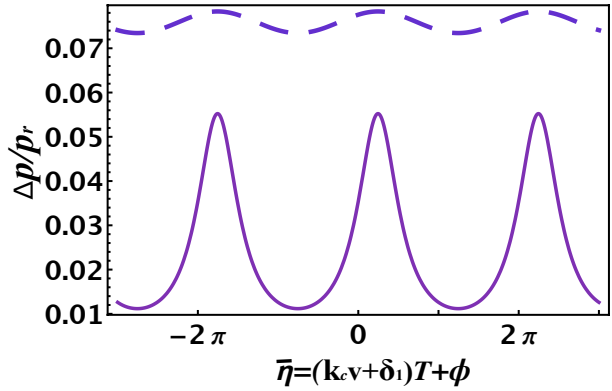}
\end{center}
\caption{ (Color online ) The dependence of the fractional momentum kick $\Delta p/(p_r)$ on the phase offset $\bar{\eta}$ at different values of  pulse repetition period $T$.  Solid purple line $T=4$ $ns$,  dashed purple line $T=50$  $ns$. The parameters of the system are: $\gamma=0.05$ $GHz$, $\Theta=\pi/4$, $\kappa=0.12$. \label{Fig:lambda}}%
\label{Fig:Setup}%
\end{figure}

In Fig.~\ref{Fig:lambda} we plot the fractional momentum kick $\Delta p/p_r $ as a function of the phase offset $\bar{\eta}$.
The radiative force (fractional momentum kick) exerted by the train of coherent pulses depends on  the atomic
velocity via Doppler shift $\bar{\eta}=(kv+\delta_1)T+\phi$. As velocity is varied across the ensemble, the maxima of the force
would occur at discrete values of velocities
\begin{equation}
v_n = ( \pi \, (2 n-\kappa)- \phi)/(k_c T) ,\, n = \ldots, -2, -1, 0, 1,2, \ldots \label{Eq:VelocityTeeth}
\end{equation}
In other words the fractional momentum kick (scattering force) spectral profile exhibits the periodic structure of the comb (see Fig.~\ref{Fig:lambda}).
As an example, for $T = 5 \, \mathrm{ns}$ and $\lambda_c=600\, \mathrm{nm}$ carrier wavelength, the force peaks are
separated by $v_{n+1}-v_n = 2\pi/(k_c T)= \lambda_c/T = 120 \, \mathrm{m/s}$ in the velocity space.
Depending on the temperature of the ensemble, the comb may have several teeth effectively interacting with the ensemble.
Notice, however, that if $\gamma T \gg 1$ (Fig.~\ref{Fig:lambda}, dashed purple line) the teeth structure of the radiative force washes out
and the atoms experience radiative force even if their velocities are far away from peaks.
In this case the power stored in the pulse is delivered to the entire ensemble.
This is in a contrast with highly-velocity selective CW laser, where the interaction
window in the velocity-space is typically $1 \, \mathrm{m/s}$.

\subsection{Maximum momentum kick}
The scattering force Eq.~(\ref{Eq:forcegeneral}) is linearly proportional to the post-pulse excited state population.
Therefore the discussion of the excited state population dependence on FC parameters in \cite{IliDer12} directly applies to the scattering force too.
In Ref. \cite{IliDer12}  we found that the maximum of $(\rho_{ee}^s)_r$ and correspondingly the maximum of fractional momentum kick for the case of equal pulse areas $\theta_1=\theta_2$ and decay rates $\gamma_1=\gamma_2$ is reached at
optimal residual detunings $\bar{\delta}_1=-\bar{\delta}_2=-mod(2\pi\kappa^{opt}, 2\pi)/T$ and optimal parameter $\kappa=\kappa^{opt}$ determined by

\begin{equation}
\kappa^{opt}=\frac1\pi\arccos(x),\label{Eq:kappaopt}
\end{equation}
where $x$ is a root of the following algebraic equation:

\begin{equation}
16 x^4 \cos ^4\frac{\Theta }{4}-32 x \cosh \frac{\gamma T }{2}\sin ^4\frac{\Theta }{4}+16 \cos \frac{\Theta }{2}-2 x^2 \left(4 \cos \frac{\Theta }{2}+3 \cos (\Theta )+9\right)=0. \label{Eq:findkappa}
\end{equation}

One can show that for the general case of non-equal decay rates, $\gamma_1\neq\gamma_2$, and pulse areas $\frac{\theta_1}{\theta_2}=\frac{\gamma_1}{\gamma_2}=\tan\chi\neq1$ and fixed  value of parameter $\kappa$,
the optimal residual detunings are determined as

\begin{equation}
\bar{\delta}_1=\left\lbrace \begin{matrix}mod(\bar{\eta}^{opt},2\pi)/T, \qquad |mod(\bar{\eta}^{opt},2\pi)|<\pi\\
(mod(\bar{\eta}^{opt},2\pi)\mp2\pi)/T, \qquad   |mod(\bar{\eta}^{opt},2\pi)|>\pi \end{matrix}  \right.,
\end{equation}

\begin{equation}
\bar{\delta}_2=\left\lbrace \begin{matrix}mod(\bar{\eta}^{opt}+2\pi\kappa,2\pi)/T, \qquad |mod(\bar{\eta}^{opt}+2\pi\kappa,2\pi)|<\pi\\
(mod(\bar{\eta}^{opt}+2\pi\kappa,2\pi)\mp2\pi)/T, \qquad   |mod(\bar{\eta}^{opt}+2\pi\kappa,2\pi)|>\pi\end{matrix}  \right. .
\end{equation}
Here

 \begin{eqnarray}
\bar{\eta}^{opt}&=&-\arctan\frac{B}A-\pi\kappa+2\pi n,\label{Eq:etta}\nonumber\\
A&=&\cos\pi\kappa( 4\cos\frac{\Theta}2(\cos2\pi\kappa-5)+(\cos\Theta+3) (3\cos2\pi\kappa+1)-16\sin^4\frac{\Theta}4 \sin^2\pi\kappa\cos(4\chi)),\nonumber\\
B&=&8\cos\left(2\chi\right)(4\sin^4\frac{\Theta}4+\sin^2\frac{\Theta}2\cos2\pi\kappa)\sin\pi\kappa.
\end{eqnarray}

At $\chi=\pi/4$ the coefficient $B$ in Eq.~(\ref{Eq:etta}) vanishes and  $\bar{\eta}^{opt}=-\kappa/2+2\pi n$, $n=0,1..$.
After substituting (\ref{Eq:etta}) into the equation for the density matrix (\ref{Eq:rgen}) one can find the optimal value of the parameter $\kappa^{opt}$, corresponding to the maximum of the post-pulse excited state population and consequently maximum fractional momentum kick.

\begin{figure}[h]
\begin{center}
\includegraphics*[width=5.2in]{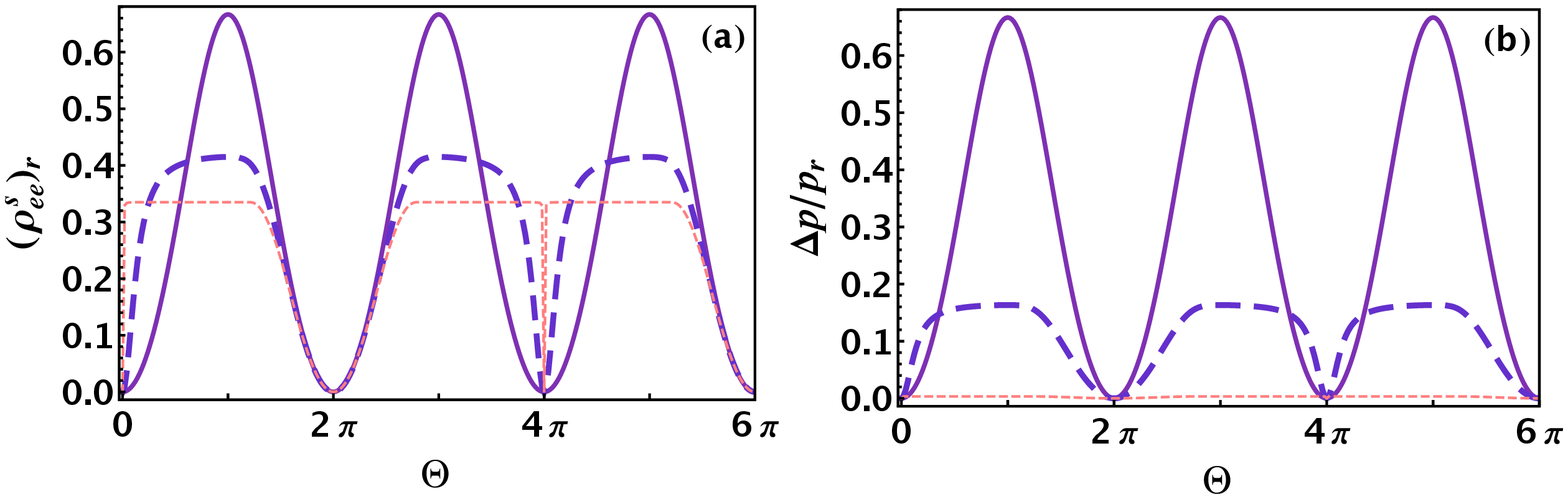}
\end{center}
\caption
{ The dependencies of the quasi-steady-state values of the post-pulse excited state population $\left(\rho_{ee}^s\right)_r$ and single pulse momentum kick $\Delta p/(p_r)$ on effective single pulse area $\Theta$ at different values of $\mu=\gamma T$: $\mu=10$ (dashed pink line),
$\mu=1/2$  (dashd blue  line),  $\mu=1/100$  (solid purple  line)
and optimal parameters $\bar{\eta}=-\pi\kappa^{opt}$,  where $\kappa^{opt}$  is obtained from Eq. (\ref{Eq:kappaopt}).
\label{Fig: maxGSSPop}} \end{figure}

The value of the single pulse area $\Theta$, maximizing the value of the fractional momentum kick is equal to $\pi+2\pi n$, $n=0,1..$.
In Fig.~\ref{Fig: maxGSSPop} (a,b) we show the dependencies of the QSS  values of the excited state population $\left(\rho_{ee}^s\right)_r$ (at $\theta_1=\theta_2$, $\gamma_1=\gamma_2$)  and corresponding single pulse momentum kick $\Delta p/p_r$ on the effective single pulse area $\Theta$.  Different curves correspond to different values of parameter $\mu=\gamma T$. The values of $\left(\rho_{ee}^s\right)_r$  and $\Delta p/ p_r$ were calculated at the
optimal value of $\kappa$, determined by Eq. (\ref{Eq:findkappa}) for each $\Theta$ and $\mu=\gamma T$.

At $\Theta=\pi$  the optimal value of the parameter $\kappa$ is equal to $1/2$, independently on the ratio of individual pulse areas $\theta_1/\theta_2$,
\begin{equation}
\kappa^{opt}_{\Theta=\pi}=\frac12.
\end{equation}
For $\kappa=1/2$ and $\Theta=\pi$ the excited state population and the fractional momentum kick are:
\begin{eqnarray}
(\rho_{ee}^s)_r(\Theta=\pi,\kappa=\frac12)&=&\frac{1}{3} e^{\gamma T /2} /\cosh(\gamma T/2),\\
\Delta p/(p_r)_{\Theta=\pi,\kappa=\pi,\bar{\eta}=-\pi/2}=\frac23\tanh(\frac{\gamma T}2).\label{Eq:PiForce}
\end{eqnarray}
The spectral resolution of scattering force at $\kappa=\kappa_{opt}$ vanishes as $\Theta\rightarrow\pi$.

As it was shown in \cite{IliDer12} the maximum post-pulse excited state population in three-level $\Lambda$-system (with $b_1=b_2=1/2$, $\theta_1=\theta_2=\sqrt{2}\pi$) is reached at $\gamma T\gg1$ and is  equal to $2/3$. Consequently the maximum of the fractional momentum kick is also $2/3$.

This result can be generalized to the case of unequal pulse areas $\theta_1\neq\theta_2$ and branching ratios $\gamma_1\neq \gamma_2$, (in case if $\theta_1\neq \pi n$, $n=0,1$). Here at $\Theta=\pi$ and $\kappa=1/2$, the  three-level $\Lambda$-system, which is initially in the ground state $|g_1\rangle$, eventually reaches the QSS with the fractional momentum kick expressed as
\begin{eqnarray}
\left(\rho_{ee}^s\right)_r=\frac{\Delta p_{max}}{p_r}=-\frac{2\sin^2(2\chi)}{(b_2- b_1) \cos (2\chi )+\cos (4\chi )-2}.\label{Eq:kickmax}
\end{eqnarray}
If the decay rates and pulse areas are $\gamma_1=\gamma\sin^2\chi=\frac{\theta_1^2}{\Theta^2}$, $\gamma_2=\gamma\cos^2\chi=\frac{\theta_2^2}{\Theta^2}$, $\theta_1=\Theta\sin^2\chi$, $\theta_2=\Theta\cos^2\chi$, the maximum fractional momentum kick (\ref{Eq:kickmax}) is equal to $2/3$. This limit is independent on the value of $\chi$ ($\theta_1\neq \pi n$ requires $\chi\neq \pi n/2$).

\subsection{Friction coefficient }
\label{Sec:FrictionCoeffBeta}

In general, one would be interested in both slowing down the atomic beam and compressing (i.e., cooling) the velocity distribution.
Cooling would occur if there is a negative velocity gradient of the radiative force $F_{sc}$.
One may introduce a friction coefficient $\beta$ by expanding  the force about some velocity $v$, corresponding to a certain value of parameter $\bar{\eta}(v)$,
\begin{equation}
  F_{sc}(v + \Delta v ) \approx F_{sc}(v) - \beta(v) \Delta v  \, .
  \label{Eq:ForceExpansionBeta}
\end{equation}
When the friction coefficient is positive  $\beta>0$, one observes the compression of velocity distribution around $v$. Negative values of $\beta$ lead to heating of the ensemble.
In the limiting case when the radiative lifetime is much shorter then the pulse repetition period there is no interference between the action of subsequent pulses on a system
and consequently no velocity dependence of the scattering force. The friction coefficient is thereby $\beta =0$ and while the ensemble slows down, there is no compression of the velocity distribution.

The friction coefficient of Eq.(\ref{Eq:ForceExpansionBeta}) may be directly determined from the analytical expression for the force
(\ref{Eq:forcegeneral}),

\begin{eqnarray}\beta&=&-16\hbar k_c^2 \sinh{\frac{\gamma T}2}\sin ^2\left(\frac{\Theta }{2}\right) \sin ^2(\pi  \kappa )\frac{B \cos \left(\bar{\eta }+\pi  \kappa \right)-A \sin \left(\bar{\eta }+\pi  \kappa \right)}{\left(A \cos \left(\bar{\eta }+\pi  \kappa \right)+B \sin \left(\bar{\eta }+\pi  \kappa \right)+C\right){}^2}\nonumber\\
C&=&8\cosh\frac{\gamma T}2 \left(\left(\cos \left(\frac{\Theta }{2}\right)+1\right) (1-\cos (2 \pi  \kappa ))+\left(1-\cos \left(\frac{\Theta
   }{2}\right)\right)^2\right),\label{Eq:bettagen}
  \end{eqnarray}
where  coefficients $A$, $B$ are defined in (\ref{Eq:etta}).

For the case of equal decay rates and  pulse areas ($\chi=\pi/4$), one has
\begin{eqnarray}
\beta(\chi=\pi/4)&=&\frac{\hbar k_c^2}{2D'}\sinh\frac{\gamma T}2\sin ^2\frac{\Theta }{2} \sin ^2(\pi  \kappa ) \cos (\pi  \kappa ) \sin (\bar{\eta}+\pi  \kappa )\times\nonumber\\ &&\left(\cos ^4\frac{\Theta }{4} \cos ^2(\pi  \kappa )-\cos \frac{\Theta
   }{2}\right).\label{Eq:bta}
\end{eqnarray}

This result depends on the effective pulse area,
$\Theta$, the product $\mu=\gamma T$  and  $\kappa=\Delta_{12}/\omega_{rep}$.
\begin{figure}[h]
\begin{center}
\includegraphics*[width=3.2in]{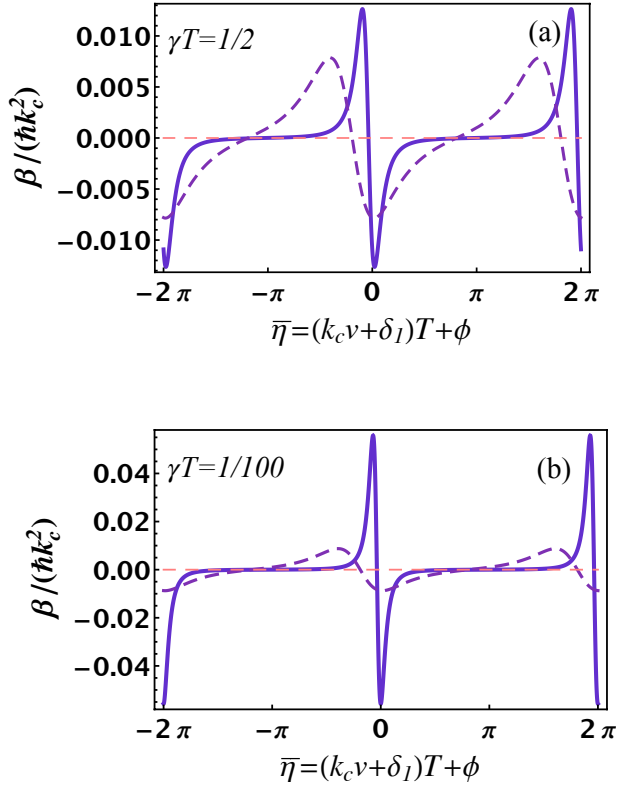}
\end{center}
\caption{(Color online)
Dependence  of the friction coefficient $\beta/\hbar k_c^2$ on
phase detuning $\bar{\eta}$ at (a)
$\gamma T=1/2$ and (b) $\gamma T=1/100$. Each panel has  three  curves with different values of pulse area $\Theta$,  $\Theta=\pi/10$ (solid purple line), $\Theta=\pi/2$ (dashed purple line),  $\Theta=\pi$ (dashed pink line).}%
\label{Fig:Beta}%
\end{figure}

In Fig.~\ref{Fig:Beta} (a,b) we plot the dependence of friction coefficient  $\beta$  (at $\theta_1=\theta_2$)  (\ref{Eq:bta}) on the phase offset $\bar{\eta}$  at different values of $\gamma T$ and $\Theta$ at  $\kappa=\kappa_{opt}$, optimally chosen for each pair of parameters $\gamma T$ and $\Theta$.
It acquires the maximum value at $\bar{\eta} = \bar{\eta}_{\beta}$,
\begin{equation}
\bar{\eta}_{\beta}=-\cos ^{-1}\left(\frac{b-\sqrt{8 a^2+b^2}}{2 a}\right)-\pi\kappa,\label{Eq:etab}
\end{equation}
where
\begin{eqnarray}
a=\cos (\pi  \kappa ) \left(\cos ^4\left(\frac{\Theta }{4}\right) \cos ^2(\pi  \kappa )-\cos \left(\frac{\Theta }{2}\right)\right),\\
b=\cosh \left(\frac{\mu }{2}\right) \left(\cos ^2\left(\frac{\Theta }{4}\right) \sin ^2(\pi  \kappa )+\sin ^4\left(\frac{\Theta
   }{4}\right)\right).
\end{eqnarray}

For the case of non-equal pulse areas $\chi\neq\pi/4$)
\begin{eqnarray}
\bar{\eta}_{\beta}&=&-\sec^{-1}\left({\frac{A^2+B^2}{A \left(C-D\right)+\sqrt{2} B \sqrt{C D-2 \left(A^2+B^2\right)-C^2}}}\right),\label{Eq:etabetagen}\nonumber\\
D&=&\sqrt{8 \left(A^2+B^2\right)+C^2},
\end{eqnarray}
where $A$, $B$ and $C$ are defined in  (\ref{Eq:etta}, \ref{Eq:bettagen}).

One can see (Fig.\,~\ref{Fig:Beta}) that as the pulse repetition rate grows (smaller $\gamma T$), smaller values of single pulse area $\Theta$ lead to larger values of the friction coefficient $\beta$.

Notice however, that at very small values of $\gamma T\ll1$  the momentum kick per pulse becomes smaller and the number of pulses needed to decelerate the atomic beam is increased.

At very large values of  $\gamma T\gg1$, while  the friction coefficient $\beta$ vanishes,  the momentum kick $\Delta p$ reaches its maximum.  If the large value of
$\gamma T\gg1$ is due to the low pulse repetition rate, then the scattering force $F_{sc}=\Delta p/T$ also becomes smaller and the overall cooling time is increased.

For $\pi$-pulse and $\theta_1=\theta_2$ ($\chi=\pi/4$) the Eq. (\ref{Eq:etab}) reduces to

\begin{equation}
\beta_{\pi}=\frac{\hbar k_c^2}2\frac{\sinh\frac{\gamma T}2 \sin ^2(\pi  \kappa ) \cos ^3(\pi  \kappa ) \sin (\bar{\eta}+\pi  \kappa )}{(\cos ^3(\pi  \kappa ) \cos (\bar{\eta}+\pi  \kappa )-(\cos (2 \pi  \kappa )-2) \cosh \left(\frac{\gamma T }{2}\right))^2}.
\end{equation}
At $\kappa=\frac12$ (chosen in order to maximize the scattering force) the friction coefficient $\beta_{\pi}$ vanishes (similar to the case when $\gamma T\gg1$). One can show that the friction coefficient at $\kappa=1/2$ and $\Theta=\pi$ turns to zero for arbitrary finite ratio of individual pulse areas $\theta_1/\theta_2$ and decay rates $\gamma_1/\gamma_2$ ($\chi\neq\pi/4$).

\subsection{Finding the optimal cooling regime}
Before discussing criteria for the optimal choice of FC parameters (single pulse area and pulse repetition rate ) we analyze the dependence of the scattering force profile on parameters $\gamma T$ and $\Theta$ at optimally chosen number of teeth $\kappa$ fitting into the energy gap between the two ground states. It is worth noticing, that the optimal value of $\kappa^{opt}$ is defined with an accuracy up the integer number, that is the values $\kappa^{opt}+n$, $n=0,1..$, where $\kappa^{opt}$ is defined from Eq. (\ref{Eq:kappaopt}), are also optimal. Analysis in this section is carried out assuming the equal individual pulse areas $\theta_1=\theta_2$ and branching ratios $b_1=b_2$ ($\chi=\pi/4$).

\begin{figure}[h]
\begin{center}
  \includegraphics*[width=3.2in]{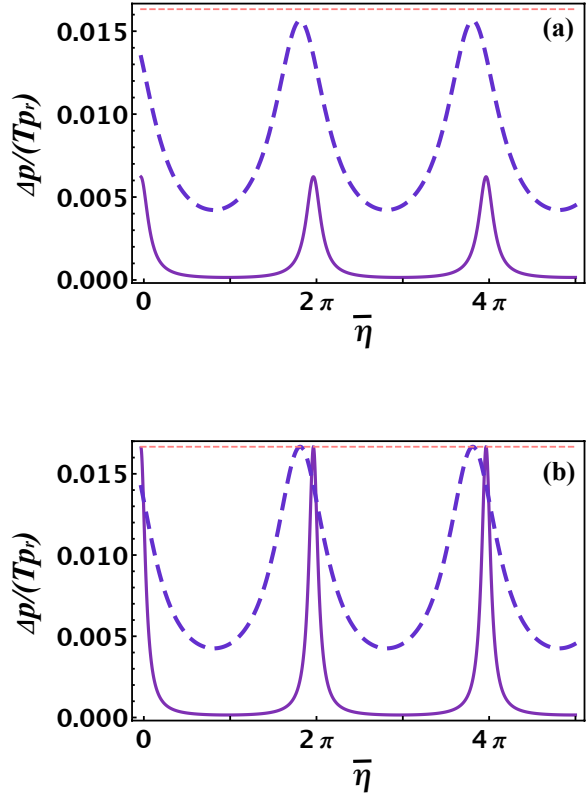}
\end{center}
\caption
{The dependence of the scattering force on the Doppler-shifted phase offset $\bar{\eta}$ at the optimal value of $\kappa=\kappa^{opt}$, chosen according to Eq. (\ref{Eq:findkappa}) at fixed value of $\mu=\gamma T$: panel (a)$\mu=1/2$, panel (b) $\mu=1/100$ and different values of
effective single pulse area $\Theta$. Different curves correspond to the distinct values of single pulse areas: $\Theta=\frac{\pi}{10}$ (solid purple line), $\Theta=\frac{\pi}{2}$ (dashed blue line), $\Theta=\pi$ (dashed pink line). \label{Fig:ForceDependence}}\end{figure}

In Fig.\,~\ref{Fig:ForceDependence} we study the dependence of the scattering force $F_{sc}$ on phase offset parameter $\bar{\eta}$ at optimally chosen $\kappa$, Eq. (\ref{Eq:kappaopt}), as the single pulse area $\Theta$ and the parameter $\mu$ vary.

One can see that at small $\mu=\gamma T$ in Fig.~\ref{Fig:ForceDependence} the maxima of the scattering force is nearly independent on the pulse area $\Theta$ as long as  $\Theta>\mu$. However, as $\Theta$ is increased the  friction coefficient becomes smaller.
As an example, at $\gamma T=1/100$ the amplitudes of scattering force corresponding to $\Theta=\pi/10$ and $\Theta=\pi/2$ are the same,  but width of the peaks is smaller at $\Theta=\pi/10$.

At higher values of parameter $\mu$ the  scattering force saturates at higher values of the pulse area $\Theta$. But the gradient of the scattering force is decreased.

At very small pulse areas $\Theta\rightarrow 0$ the scattering force vanishes (as well as the momentum kick $\Delta p$) regardless of the parameter $\gamma T$.

To summarize,  at lower pulse repetition rates $\gamma T\gg1$ and larger values of pulse area one can obtain larger momentum kick and smaller scattering force and compression rate.
At larger pulse repetition rate ($\gamma T\ll1$) and properly chosen $\Theta$ one can obtain the maximum of compression rate, but smaller momentum kick.
In the first case the cooling time is increased and the compression of the velocity distribution is slow.
In the second case the number of pulses needed to decelerate the beam is increased and the scattering force velocity capture range is decreased.

To find the opimal cooling regime one has to compromise between the fast slowing of the entire ensemble and its velocity distribution compression rate.
In case when the scattering force rapidly vanishes in the vicinity of its maxima only the atoms within narrow groups of velocity are decelerated.
Below we show that this problem can be mitigated.
The spectral dependence of the scattering force can be varied in time,
so that the positions of maxima follow the center of velocity distribution of the decelerating ensemble.
In this case those atoms, which initially were outside of the scattering force velocity capture range and not decelerated will be eventually captured by the force profile being moved in the spectral domain (e.g. by changing the CEPO  $\phi$).

However, if the initial atomic beam is too fast and (or) the velocity distribution is too wide, one can be interested initially in slowing down the ensemble, so that the cooling distance will be not too large.
In this case one would prefer to have a broad scattering force profile (wide velocity capture range). The amplitude of the scattering force has to be large enough to mitigate the increase of cooling time and consequently cooling distance. This can be realized at larger pulse areas $\Theta$.
For example, at $\mu\sim 1/2$, $\Theta\sim\pi/2$ for the atoms with the excited state lifetime $\tau\sim 15$ $ns$ at the laser field wavelength $\lambda=589$ $nm$, the velocity capture range $\Delta v_{cptr}$ is estimated by $\Delta v_{cptr}\sim 20$ $m/s$.  The scattering force amplitude     is quite the same as its maximum value, reached at $\Theta=\pi$.

At $\Theta=\pi$ the velocity capture range can be extended up to $\lambda/\tau_p$, where $\tau_p$ is the duration of pulse. For $\tau_p\sim 1 $  $ps$ and $\lambda=589$ $nm$ the maximum velocity capture range is very broad $\Delta v_{cptr}^{max}\sim 5.89\times 10^5$ $m/s$.

If the intial velocity distribution is already narrow and (or) the central velocity value is not too high, the priority can be given to the
fast velocity distribution compression.

The optimal set values of pulse area and pulse repetition rate can be chosen based on the initial velocity distribution,  desired velocity compression rate and the limiting factors such as given cooling length and the laser power.


\section{Evolution of the velocity distribution}
\label{Ssec:slowing}

Now we turn to the dynamics of slowing down and cooling an entire atomic ensemble,
characterized by some velocity distribution $f(v,t)$ (time-dependence is caused by radiative force).

\subsection{No-cooling theorem for fixed FC parameters}

Suppose that the positions of FC teeth remain fixed in the frequency domain during the deceleration.
As the atoms slow down, they come in and out of resonances with different FC teeth. 
The gradient of the scattering force changes its sign (see Fig.~\ref{Fig:Beta}) as the Doppler-shifted phase (velocity) vary.  As a result the sustained cooling can not be realized if the positions of FC teeth remain fixed in frequency space during deceleration.  This can be demonstrated as follows.
Suppose the parameters of the frequency comb remain fixed. As a result of scattering $N$ pulses the atom with initial velocity $v_i$ will be decelerated to the final velocity $v_f$ determined from the implicit equation
\begin{eqnarray}
N v_r&=&\frac{2 \csc ^2\frac{\Theta }{2}\csc ^2\pi  \kappa }{ k_c T \sinh\frac{\gamma T}2}\left(  k_c T(v_f-v_i) \cosh\frac{\gamma T }{2}\left(\cos ^4\frac{\Theta }{4} \sin ^2\pi  \kappa+\sin ^4\frac{\Theta
   }{4}\right)+\right. \nonumber \\ &&
  \left. \cos \pi  \kappa  \left(\cos ^4\frac{\Theta }{4}\cos ^2\pi  \kappa -\cos \frac{\Theta }{2}\right) (\sin (k_cTv_f+\pi  \kappa )-\sin (k_cTv_i+\pi  \kappa ))\right).\qquad
\label{Eq:FixedCombDeltaV}
\end{eqnarray}
where $v_r \equiv p_r/M$ is the recoil velocity. This equation was obtained by integrating Eq. (\ref{Eq:scforce}).

Eq.~(\ref{Eq:FixedCombDeltaV}) implies that
the decrement in velocities would vary across the ensemble.
Yet if we fix the change of velocity equal to the spacing between the teeth,
$v_f=v_i-\lambda_c/T$, we find that the required number of pulses $N_0$ (or time $N_0 T$),
\begin{equation}\label{N0}
N_0 = \frac{2\lambda_c  \csc ^2\frac{\Theta }{2}\csc ^2\pi  \kappa }{T  v_r \sinh\frac{\gamma T}2} \cosh\frac{\gamma T }{2}\left(\cos ^4\frac{\Theta }{4}\sin ^2(\pi  \kappa )+\sin ^4\frac{\Theta
   }{4}\right),
\end{equation}
does not dependent on the initial value $v_i$. This implies that if we start with a certain
velocity distribution $f(v)$,
the entire distribution is uniformly shifted by $-\lambda_c/T$ every $N_0$ pulses:
$f(v) \rightarrow_{N_0} f(v+\lambda_c/T)$.
Thus, the radiative force exerted by FC with fixed parameters does not lead to velocity compression
--- there is no cooling.


Notice that the above analysis has neglected variation of intensity across comb teeth. Also
while there is no compression of the velocity distribution, there is a residual heating due
to atomic recoil (this arises from treatments beyond our model,
see, e.g., Ref.~\cite{MetStr99Book}).

\subsection{Cooling via tuning the FC}
In order to compress the velocity distribution, one has to maintain the positive gradient of the absolute value of the scattering force in the vicinity of the center of velocity distribution. To attain this condition, the scattering force profile has to follow the center of velocity distribution,  moving towards the smaller velocities (frequencies) during the process of deceleration.  In other words,  the FC tooth closest  to the atomic transition frequency $\omega_{eg_1}$   (in the reference frame moving with the center of velocity distribution) has to be somewhat red-detuned from  $\omega_{eg_1}$.  Tuning the positions of the FC teeth and consequently the scattering force profile can be achieved by tuning the phase of pulses during the cooling process \cite{IliAhmDer11} .

Initially, we start with some velocity distribution $f(v,t=0)$.
To optimize the number of cooled atoms, we focus on  atoms with
velocities grouped around
the position of the maximum of $f(v,t=0)$, i.e. the most probable velocity $v_{mp}(t=0)$.
Radiative force will cause both the distribution $f(v,t)$ and the
most probable velocity $v_{mp}(t)$ to evolve  in time.

To maximize the rate of compression, the friction coefficient needs to be kept at
its maximum value at $v_{mp}(t)$.
We may satisfy this requirement
by tuning the phase offset $\phi(mT)=\Phi((m+1)T)-\Phi(mT)$ according to
\begin{equation}
\phi(t) = -\left( \delta + k_c v_{mp}(t) \right) T - \bar{\eta}_{\beta} \, ,
\end{equation}
where $\bar{\eta}_{\beta}$, Eq.~(\ref{Eq:etabetagen}),
depends only on  (time-independent) values of $\gamma T$,  $\Theta$ and $\Delta_{12}/\omega_{rep}$. As $v_{mp}(t)$ becomes smaller
due to the radiative force, the offset phase needs to be reduced.

We may find  required pulse-to-pulse increment of the phase offset explicitly
\begin{equation}
\Delta \phi_T=\phi((m+1)T)-\phi(mT)=-\frac{k_cT}{M}\Delta p(\bar{\eta}_{\beta})\,.\label{Eq:DeltaPhi}%
\end{equation}

When the phase offset is driven according to (\ref{Eq:DeltaPhi}), there is a dramatic
change in time-evolution of velocities of individual atoms.
As the phase offset is varied over time,
the entire frequency-comb structure shifts towards lower frequencies.
As the teeth sweep through the velocity space, atomic $v(t)$ trajectories are ``snow-plowed''
by teeth, ultimately leading to narrow velocity spikes collected on the teeth.
This emergence of ``velocity comb'' was discussed in Ref. \cite{IliAhmDer11} for two-level system.
Formally, we may separate initial velocities into groups

\begin{equation}
v_{mp}(t=0)+\left( 2\pi (n-1)-\bar{\eta}_{\beta}  \right)/k_c T
< v(t=0) < v_{mp}(t=0)+\left(2\pi n- \bar{\eta}_{\beta}   \right)/k_c T,  n=0,\pm 1 \ldots
\end{equation}
The width of each velocity group is equal to the distance between neighboring teeth in
velocity space, $2\pi/k_c T$.  As a result of ``snow-plowing'', the $n^\mathrm{th}$ group
will be piled up at $v_n(t)=v_{mp}(t)+ 2\pi n /k_c T$. The final velocity spread of individual
velocity groups will be limited by the Doppler temperature, $T_D = \hbar \gamma/2 k_B$.

\begin{figure}[h]
\begin{center}
\includegraphics*[scale=0.5]{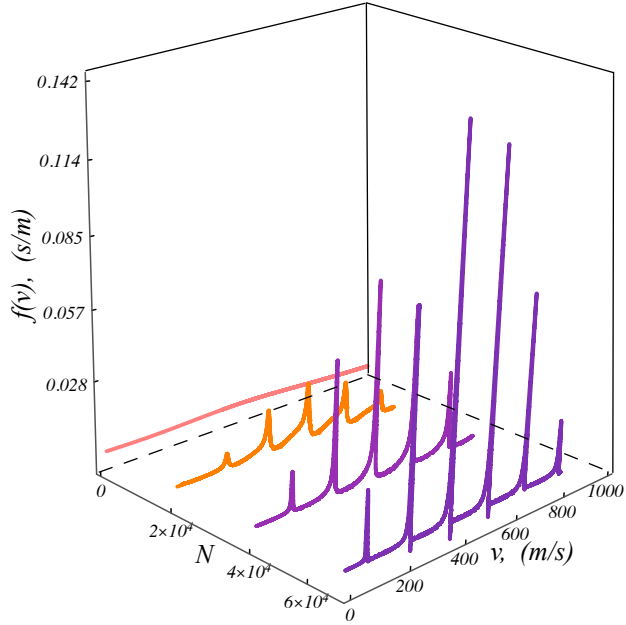}
\end{center}
\caption{Time-evolution of velocity distribution for a thermal beam subjected to
a coherent train of laser pulses. Pulse-to-pulse phase offset of the train is varied
linearly in time  as prescribed by Eq.~(\ref{Eq:DeltaPhi}).  $N$ is the number of pulses.   (a) Atomic and pulse train parameters are:
$\gamma T=0.25$, $\Theta=\pi/2$. The optimal phase detuning is $\bar{\eta}=-1.23$.
The center of initial velocity distribution is $v_{mp}=500$ m/s.\label{Fig:fvt}}
\end{figure}

To illustrate the train-driven time-evolution for the entire ensemble,
we consider a 1D thermal beam characterized by the initial velocity distribution
\begin{equation}
f(v,t=0)=\frac{v^3}{2\tilde{v_0}^4}\exp(-\frac{v^2}{2\tilde{v_0}^2}).
\end{equation}
The most probable $v_{mp}$, average $v_{ave}$ and the r.m.s. $v_{rms}$ values are expressed in terms of $\widetilde{v}_0$ as
\begin{eqnarray}
v_{mp}=\sqrt{3}\widetilde{v}_0,\nonumber\\
v_{ave}=\sqrt{\frac{9\pi}{8}}\widetilde{v}_0,\nonumber\\
v_{rms}=2\widetilde{v}_0.
\end{eqnarray}
$v_{mp}$ is the most probable velocity at $t=0$.
A typical time-evolution of the velocity distribution is shown in Fig.~\ref{Fig:fvt}.  Local compression of velocity distribution happens near
the points $v_c(t)+\lambda_c n/T$, $n=0,\pm1..$,
where $v_c$ is the time-dependent position of  velocity distribution center.
Clearly, velocity distribution, while initially being continuous, after a certain number
of pulses develops a comb-like profile.
This  is the ``velocity comb'' of sharp peaks separated by equal intervals $\lambda_c/T$ in the velocity space.

\section{Conclusion}

In this paper we studied Doppler cooling of a three-level $\Lambda$-type system driven by a train of ultra-short laser pulses. Analytical expression for the scattering force was obtained and its dependence on the FC parameters was analyzed.
The scattering force $F_{sc}$ is linearly proportional to the quasi-steady-state post-pulse excited state population.
Its spectral (velocity) dependence exhibits periodic pattern mimicking the spectrum of the frequency comb.
The contrast of the spectral profile of $F_{sc}$ is a function of the ratio between the excited state lifetime and the pulse repetition period, the effective single pulse area and the residual detunings $\bar{\delta_j}$ between the frequencies of individual transitions and nearest FC teeth.
In a particular case when the pulse repetition period  is much longer than the lifetime of the excited state, the spectral dependence of the  scattering force reflects the broad-band spectral profile of a singe pulse.

The residual detunings $\bar{\delta}_j$ can be optimized to maximize the scattering force.  At optimally chosen detunings the maximum of the scattering force is reached at single pulse area equal to $\pi$. However for $\pi$-pulses the spectral dependence of the scattering force is lost and consequently the friction coefficient vanishes.  To optimize the cooling process one has to compromise between maximizing the scattering force and its velocity capture range and maintaining the sufficient gradient of the scattering force (friction coefficient).
The spectral profile of the scattering force and consequently the friction coefficient can be varied in time to follow the moving center of the velocity distribution of decelerating ensemble.  This can be realized by simply tuning the carrier envelope phase offset. Such manipulation enables sustained velocity distribution compression as the atoms slow down.
As a result, initially smooth velocity distribution of a thermal beam evolves into a series of narrow groups of velocities separated by $\lambda_c/T$,
so called ``velocity comb''.


\section*{Acknowledgments}
This work was supported in part by the NSF and ARO. We would like to thank Mahmoud Ahmad for discussions.

\begin{thebibliography}{30}
\expandafter\ifx\csname natexlab\endcsname\relax\def\natexlab#1{#1}\fi
\expandafter\ifx\csname bibnamefont\endcsname\relax
  \def\bibnamefont#1{#1}\fi
\expandafter\ifx\csname bibfnamefont\endcsname\relax
  \def\bibfnamefont#1{#1}\fi
\expandafter\ifx\csname citenamefont\endcsname\relax
  \def\citenamefont#1{#1}\fi
\expandafter\ifx\csname url\endcsname\relax
  \def\url#1{\texttt{#1}}\fi
\expandafter\ifx\csname urlprefix\endcsname\relax\def\urlprefix{URL }\fi
\providecommand{\bibinfo}[2]{#2}
\providecommand{\eprint}[2][]{\url{#2}}

\bibitem[{\citenamefont{Hansch and Schawlow}(1975)}]{HanSch75}
\bibinfo{author}{\bibfnamefont{T.}~\bibnamefont{Hansch}} \bibnamefont{and}
  \bibinfo{author}{\bibfnamefont{A.}~\bibnamefont{Schawlow}},
  \bibinfo{journal}{Opt. Comm.} \textbf{\bibinfo{volume}{13}},
  \bibinfo{pages}{68} (\bibinfo{year}{1975}).

\bibitem[{\citenamefont{Minogin and Letokhov}(1987)}]{MinLet87Book}
\bibinfo{author}{\bibfnamefont{V.}~\bibnamefont{Minogin}} \bibnamefont{and}
  \bibinfo{author}{\bibfnamefont{V.}~\bibnamefont{Letokhov}},
  \emph{\bibinfo{title}{Laser Light Pressure on Atoms}}
  (\bibinfo{publisher}{Gordon and Breach}, \bibinfo{address}{New York},
  \bibinfo{year}{1987}).

\bibitem[{\citenamefont{Metcalf and van~der Straten}(1999)}]{MetStr99Book}
\bibinfo{author}{\bibfnamefont{H.~J.} \bibnamefont{Metcalf}} \bibnamefont{and}
  \bibinfo{author}{\bibfnamefont{P.}~\bibnamefont{van~der Straten}},
  \emph{\bibinfo{title}{Laser Cooling and Trapping}}
  (\bibinfo{publisher}{Springer,New York}, \bibinfo{year}{1999}).

\bibitem[{\citenamefont{Berman and Malinovsky}(2010)}]{BerMal10_Book}
\bibinfo{author}{\bibfnamefont{P.~R.} \bibnamefont{Berman}} \bibnamefont{and}
  \bibinfo{author}{\bibfnamefont{V.~S.} \bibnamefont{Malinovsky}},
  \emph{\bibinfo{title}{Principles of Laser Spectroscopy and Quantum Optics}}
  (\bibinfo{publisher}{Princeton University Press}, \bibinfo{year}{2010}).

\bibitem[{\citenamefont{Hoffnagle}(1988)}]{Hof88}
\bibinfo{author}{\bibfnamefont{J.}~\bibnamefont{Hoffnagle}},
  \bibinfo{journal}{J. Opt. Lett.} \textbf{\bibinfo{volume}{13}},
  \bibinfo{pages}{102} (\bibinfo{year}{1988}).

\bibitem[{\citenamefont{Strohmeier et~al.}(1989)}]{Str89}
\bibinfo{author}{\bibnamefont{Strohmeier}} \bibnamefont{et~al.},
  \bibinfo{journal}{Opt. Comm.} \textbf{\bibinfo{volume}{73}},
  \bibinfo{pages}{451} (\bibinfo{year}{1989}).

\bibitem[{\citenamefont{Watanabe et~al.}(1996)\citenamefont{Watanabe, Ohmukai,
  Tanaka, Hayasaka, Imajo, and Urabe}}]{WatOhmTan96}
\bibinfo{author}{\bibfnamefont{M.}~\bibnamefont{Watanabe}},
  \bibinfo{author}{\bibfnamefont{R.}~\bibnamefont{Ohmukai}},
  \bibinfo{author}{\bibfnamefont{U.}~\bibnamefont{Tanaka}},
  \bibinfo{author}{\bibfnamefont{K.}~\bibnamefont{Hayasaka}},
  \bibinfo{author}{\bibfnamefont{H.}~\bibnamefont{Imajo}}, \bibnamefont{and}
  \bibinfo{author}{\bibfnamefont{S.}~\bibnamefont{Urabe}}, \bibinfo{journal}{J.
  Opt. Soc. Am. B} \textbf{\bibinfo{volume}{13}}, \bibinfo{pages}{2377}
  (\bibinfo{year}{1996}).

\bibitem[{\citenamefont{Kielpinski}(2006)}]{Kie06}
\bibinfo{author}{\bibfnamefont{D.}~\bibnamefont{Kielpinski}},
  \bibinfo{journal}{Phys. Rev. A} \textbf{\bibinfo{volume}{73}},
  \bibinfo{pages}{063407} (\bibinfo{year}{2006}).

\bibitem[{\citenamefont{Ilinova et~al.}(2011)\citenamefont{Ilinova, Ahmad, and
  Derevianko}}]{IliAhmDer11}
\bibinfo{author}{\bibfnamefont{E.}~\bibnamefont{Ilinova}},
  \bibinfo{author}{\bibfnamefont{M.}~\bibnamefont{Ahmad}}, \bibnamefont{and}
  \bibinfo{author}{\bibfnamefont{A.}~\bibnamefont{Derevianko}},
  \bibinfo{journal}{ArXiv e-prints}  (\bibinfo{year}{2011}),
  \eprint{1105.0665}.

\bibitem[{\citenamefont{Kazantsev}(1974)}]{Kaz74}
\bibinfo{author}{\bibfnamefont{A.}~\bibnamefont{Kazantsev}},
  \bibinfo{journal}{Zh. Exp. Theor. Fiz.} \textbf{\bibinfo{volume}{66}},
  \bibinfo{pages}{1599} (\bibinfo{year}{1974}).

\bibitem[{\citenamefont{Noelle et~al.}(1996)\citenamefont{Noelle, Noelle,
  Schmand, and Andra}}]{NoeNoeSch96}
\bibinfo{author}{\bibfnamefont{B.}~\bibnamefont{Noelle}},
  \bibinfo{author}{\bibfnamefont{H.}~\bibnamefont{Noelle}},
  \bibinfo{author}{\bibfnamefont{J.}~\bibnamefont{Schmand}}, \bibnamefont{and}
  \bibinfo{author}{\bibfnamefont{H.~J.} \bibnamefont{Andra}},
  \bibinfo{journal}{Europhys. Lett.} \textbf{\bibinfo{volume}{33}},
  \bibinfo{pages}{261} (\bibinfo{year}{1996}).

\bibitem[{\citenamefont{Goepfert et~al.}(1997)\citenamefont{Goepfert, Bloch,
  Haubrich, Lison, Sch\"utze, Wynands, and Meschede}}]{GoeBloHau97}
\bibinfo{author}{\bibfnamefont{A.}~\bibnamefont{Goepfert}},
  \bibinfo{author}{\bibfnamefont{I.}~\bibnamefont{Bloch}},
  \bibinfo{author}{\bibfnamefont{D.}~\bibnamefont{Haubrich}},
  \bibinfo{author}{\bibfnamefont{F.}~\bibnamefont{Lison}},
  \bibinfo{author}{\bibfnamefont{R.}~\bibnamefont{Sch\"utze}},
  \bibinfo{author}{\bibfnamefont{R.}~\bibnamefont{Wynands}}, \bibnamefont{and}
  \bibinfo{author}{\bibfnamefont{D.}~\bibnamefont{Meschede}},
  \bibinfo{journal}{Phys. Rev. A} \textbf{\bibinfo{volume}{56}},
  \bibinfo{pages}{R3354} (\bibinfo{year}{1997}),
  \urlprefix\url{http://link.aps.org/doi/10.1103/PhysRevA.56.R3354}.

\bibitem[{\citenamefont{S\"oding et~al.}(1997)\citenamefont{S\"oding, Grimm,
  Ovchinnikov, Bouyer, and Salomon}}]{SodGriOvc97}
\bibinfo{author}{\bibfnamefont{J.}~\bibnamefont{S\"oding}},
  \bibinfo{author}{\bibfnamefont{R.}~\bibnamefont{Grimm}},
  \bibinfo{author}{\bibfnamefont{Y.~B.} \bibnamefont{Ovchinnikov}},
  \bibinfo{author}{\bibfnamefont{P.}~\bibnamefont{Bouyer}}, \bibnamefont{and}
  \bibinfo{author}{\bibfnamefont{C.}~\bibnamefont{Salomon}},
  \bibinfo{journal}{Phys. Rev. Lett.} \textbf{\bibinfo{volume}{78}},
  \bibinfo{pages}{1420} (\bibinfo{year}{1997}),
  \urlprefix\url{http://link.aps.org/doi/10.1103/PhysRevLett.78.1420}.

\bibitem[{\citenamefont{Schibli et~al.}(2008)\citenamefont{Schibli, Hartl,
  Yost, Martin, Marcinkevicius, Fermann, and Ye}}]{SchHarYos08}
\bibinfo{author}{\bibfnamefont{T.~R.} \bibnamefont{Schibli}},
  \bibinfo{author}{\bibfnamefont{I.}~\bibnamefont{Hartl}},
  \bibinfo{author}{\bibfnamefont{D.~C.} \bibnamefont{Yost}},
  \bibinfo{author}{\bibfnamefont{M.~J.} \bibnamefont{Martin}},
  \bibinfo{author}{\bibfnamefont{A.}~\bibnamefont{Marcinkevicius}},
  \bibinfo{author}{\bibfnamefont{M.~E.} \bibnamefont{Fermann}},
  \bibnamefont{and} \bibinfo{author}{\bibfnamefont{J.}~\bibnamefont{Ye}},
  \bibinfo{journal}{Nat. Photon.} \textbf{\bibinfo{volume}{2}},
  \bibinfo{pages}{355} (\bibinfo{year}{2008}).

\bibitem[{\citenamefont{Adler et~al.}(2009)\citenamefont{Adler, Cossel, Thorpe,
  Hartl, Fermann, and Ye}}]{AdlCosTho09}
\bibinfo{author}{\bibfnamefont{F.}~\bibnamefont{Adler}},
  \bibinfo{author}{\bibfnamefont{K.~C.} \bibnamefont{Cossel}},
  \bibinfo{author}{\bibfnamefont{M.~J.} \bibnamefont{Thorpe}},
  \bibinfo{author}{\bibfnamefont{I.}~\bibnamefont{Hartl}},
  \bibinfo{author}{\bibfnamefont{M.~E.} \bibnamefont{Fermann}},
  \bibnamefont{and} \bibinfo{author}{\bibfnamefont{J.}~\bibnamefont{Ye}},
  \bibinfo{journal}{Opt. Lett.} \textbf{\bibinfo{volume}{34}},
  \bibinfo{pages}{1330} (\bibinfo{year}{2009}).

\bibitem[{\citenamefont{Vodopyanov et~al.}(2011)\citenamefont{Vodopyanov,
  Sorokin, Sorokina, and Schunemann}}]{VodSorSor11}
\bibinfo{author}{\bibfnamefont{K.}~\bibnamefont{Vodopyanov}},
  \bibinfo{author}{\bibfnamefont{E.}~\bibnamefont{Sorokin}},
  \bibinfo{author}{\bibfnamefont{I.~T.} \bibnamefont{Sorokina}},
  \bibnamefont{and} \bibinfo{author}{\bibfnamefont{P.~G.}
  \bibnamefont{Schunemann}}, \bibinfo{journal}{Opt. Lett.}
  \textbf{\bibinfo{volume}{36}}, \bibinfo{pages}{2275} (\bibinfo{year}{2011}).

\bibitem[{\citenamefont{Marian et~al.}(2004)\citenamefont{Marian, Stowe,
  Lawall, Felinto, and Ye}}]{MarStoLaw04}
\bibinfo{author}{\bibfnamefont{A.}~\bibnamefont{Marian}},
  \bibinfo{author}{\bibfnamefont{M.~C.} \bibnamefont{Stowe}},
  \bibinfo{author}{\bibfnamefont{J.~R.} \bibnamefont{Lawall}},
  \bibinfo{author}{\bibfnamefont{D.}~\bibnamefont{Felinto}}, \bibnamefont{and}
  \bibinfo{author}{\bibfnamefont{J.}~\bibnamefont{Ye}},
  \bibinfo{journal}{Science} \textbf{\bibinfo{volume}{17}},
  \bibinfo{pages}{2063} (\bibinfo{year}{2004}).

\bibitem[{\citenamefont{Prudnikov and Arimondo}(2003)}]{PruAri03}
\bibinfo{author}{\bibfnamefont{O.~N.} \bibnamefont{Prudnikov}}
  \bibnamefont{and} \bibinfo{author}{\bibfnamefont{E.}~\bibnamefont{Arimondo}},
  \bibinfo{journal}{J. Opt. Soc. Am. B} \textbf{\bibinfo{volume}{20}},
  \bibinfo{pages}{909} (\bibinfo{year}{2003}).

\bibitem[{\citenamefont{Aspect et~al.}(1988)\citenamefont{Aspect, Arimondo,
  Kaiser, Vansteenkiste, and Cohen-Tannoudji}}]{AspAriKai88}
\bibinfo{author}{\bibfnamefont{A.}~\bibnamefont{Aspect}},
  \bibinfo{author}{\bibfnamefont{E.}~\bibnamefont{Arimondo}},
  \bibinfo{author}{\bibfnamefont{R.}~\bibnamefont{Kaiser}},
  \bibinfo{author}{\bibfnamefont{N.}~\bibnamefont{Vansteenkiste}},
  \bibnamefont{and}
  \bibinfo{author}{\bibfnamefont{C.}~\bibnamefont{Cohen-Tannoudji}},
  \bibinfo{journal}{Phys. Rev. Lett.} \textbf{\bibinfo{volume}{61}},
  \bibinfo{pages}{826} (\bibinfo{year}{1988}),
  \urlprefix\url{http://link.aps.org/doi/10.1103/PhysRevLett.61.826}.

\bibitem[{\citenamefont{Kasevich and Chu}(1992)}]{KasChu92}
\bibinfo{author}{\bibfnamefont{M.}~\bibnamefont{Kasevich}} \bibnamefont{and}
  \bibinfo{author}{\bibfnamefont{S.}~\bibnamefont{Chu}},
  \bibinfo{journal}{Phys. Rev. Lett.} \textbf{\bibinfo{volume}{69}},
  \bibinfo{pages}{1741} (\bibinfo{year}{1992}),
  \urlprefix\url{http://link.aps.org/doi/10.1103/PhysRevLett.69.1741}.

\bibitem[{\citenamefont{Gupta et~al.}(1993)\citenamefont{Gupta, Xie, Padua,
  Batelaan, and Metcalf}}]{GupXiePad93}
\bibinfo{author}{\bibfnamefont{R.}~\bibnamefont{Gupta}},
  \bibinfo{author}{\bibfnamefont{C.}~\bibnamefont{Xie}},
  \bibinfo{author}{\bibfnamefont{S.}~\bibnamefont{Padua}},
  \bibinfo{author}{\bibfnamefont{H.}~\bibnamefont{Batelaan}}, \bibnamefont{and}
  \bibinfo{author}{\bibfnamefont{H.}~\bibnamefont{Metcalf}},
  \bibinfo{journal}{Phys. Rev. Lett.} \textbf{\bibinfo{volume}{71}},
  \bibinfo{pages}{3087} (\bibinfo{year}{1993}).

\bibitem[{\citenamefont{Aspect et~al.}(1986)\citenamefont{Aspect, Dalibard,
  Heidmann, Salomon, and Cohen-Tannoudji}}]{AspDalHei86}
\bibinfo{author}{\bibfnamefont{A.}~\bibnamefont{Aspect}},
  \bibinfo{author}{\bibfnamefont{J.}~\bibnamefont{Dalibard}},
  \bibinfo{author}{\bibfnamefont{A.}~\bibnamefont{Heidmann}},
  \bibinfo{author}{\bibfnamefont{C.}~\bibnamefont{Salomon}}, \bibnamefont{and}
  \bibinfo{author}{\bibfnamefont{C.}~\bibnamefont{Cohen-Tannoudji}},
  \bibinfo{journal}{Phys. Rev. Lett.} \textbf{\bibinfo{volume}{57}},
  \bibinfo{pages}{1688} (\bibinfo{year}{1986}),
  \urlprefix\url{http://link.aps.org/doi/10.1103/PhysRevLett.57.1688}.

\bibitem[{\citenamefont{Zhu et~al.}(1991)\citenamefont{Zhu, Oates, and
  Hall}}]{ZhuOatHal91}
\bibinfo{author}{\bibfnamefont{M.}~\bibnamefont{Zhu}},
  \bibinfo{author}{\bibfnamefont{C.~W.} \bibnamefont{Oates}}, \bibnamefont{and}
  \bibinfo{author}{\bibfnamefont{J.~L.} \bibnamefont{Hall}},
  \bibinfo{journal}{Phys. Rev. Lett.} \textbf{\bibinfo{volume}{67}},
  \bibinfo{pages}{46} (\bibinfo{year}{1991}).

\bibitem[{\citenamefont{Schr\"oder et~al.}(1990)}]{SchKleBoo90etal}
\bibinfo{author}{\bibfnamefont{S.}~\bibnamefont{Schr\"oder}}
  \bibnamefont{et~al.}, \bibinfo{journal}{Phys. Rev. Lett.}
  \textbf{\bibinfo{volume}{64}}, \bibinfo{pages}{2901} (\bibinfo{year}{1990}).

\bibitem[{\citenamefont{Miesner et~al.}(1996)}]{MieGriGri96etal}
\bibinfo{author}{\bibfnamefont{H.-J.} \bibnamefont{Miesner}}
  \bibnamefont{et~al.}, \bibinfo{journal}{Phys. Rev. Lett.}
  \textbf{\bibinfo{volume}{77}}, \bibinfo{pages}{623} (\bibinfo{year}{1996}).

\bibitem[{\citenamefont{Ilinova and Derevianko}(2012)}]{IliDer12}
\bibinfo{author}{\bibfnamefont{E.}~\bibnamefont{Ilinova}} \bibnamefont{and}
  \bibinfo{author}{\bibfnamefont{A.}~\bibnamefont{Derevianko}},
  \bibinfo{journal}{arXiv:1203.0034}  (\bibinfo{year}{2012}).

\bibitem[{\citenamefont{Harris}(1997)}]{Har97}
\bibinfo{author}{\bibfnamefont{S.~E.} \bibnamefont{Harris}},
  \bibinfo{journal}{Physics Today} \textbf{\bibinfo{volume}{50}},
  \bibinfo{pages}{36} (\bibinfo{year}{1997}).

\bibitem[{\citenamefont{Moreno and Vianna}(2011)}]{MorVia11}
\bibinfo{author}{\bibfnamefont{M.~P.} \bibnamefont{Moreno}} \bibnamefont{and}
  \bibinfo{author}{\bibfnamefont{S.~S.} \bibnamefont{Vianna}},
  \bibinfo{journal}{J. Opt. Soc. Am. B} \textbf{\bibinfo{volume}{28}},
  \bibinfo{pages}{1124} (\bibinfo{year}{2011}).

\bibitem[{\citenamefont{Soares and de~Araujo}(2007)}]{SoaAra07}
\bibinfo{author}{\bibfnamefont{A.~A.} \bibnamefont{Soares}} \bibnamefont{and}
  \bibinfo{author}{\bibfnamefont{L.~E.~E.} \bibnamefont{de~Araujo}},
  \bibinfo{journal}{Phys. Rev. A} \textbf{\bibinfo{volume}{76}},
  \bibinfo{pages}{043818} (\bibinfo{year}{2007}),
  \urlprefix\url{http://link.aps.org/doi/10.1103/PhysRevA.76.043818}.

\bibitem[{\citenamefont{Soares and Araujo}(2010)}]{SoaAra10}
\bibinfo{author}{\bibfnamefont{A.}~\bibnamefont{Soares}} \bibnamefont{and}
  \bibinfo{author}{\bibfnamefont{E.~E.} \bibnamefont{Araujo}},
  \bibinfo{journal}{J. Phys. B} \textbf{\bibinfo{volume}{43}},
  \bibinfo{pages}{085003} (\bibinfo{year}{2010}).

\end{thebibliography}

\end{document}